\documentclass[reprint]{revtex4-1}
\usepackage[english]{babel}
\usepackage{graphicx}
\usepackage[font={small},justification=centerlast]{caption}
\usepackage{amsmath}
\usepackage{float}
\usepackage{wrapfig}
\usepackage{bm}
\usepackage{mathtools}
\usepackage{xr}
\usepackage{amssymb}
\usepackage{color}
\usepackage{epstopdf}

\draft 

\usepackage{lipsum}

\begin{document}


\title{Interparticle torques suppress motility-induced phase separation for rodlike particles}



\author{Robin van Damme$^1$}
\author{Jeroen Rodenburg$^2$}
\author{Ren\'e van Roij$^2$}
\author{Marjolein Dijkstra$^1$}
\affiliation{${}^1$ Soft Condensed Matter, Debye Institute for Nanomaterials Science, Utrecht University, Princetonplein 1, 3584 CC Utrecht, The Netherlands\\
${}^2$ Institute for Theoretical Physics, Utrecht University, Princetonplein 5, 3584 CC Utrecht, The Netherlands
}

\date{\today}

\begin{abstract}
To study the role of torque in motility-induced phase separation (MIPS), we simulate a system of self-propelled particles whose shape varies smoothly from isotropic (disks/spheres) to weakly elongated (rods). We construct the phase diagrams of 2D active disks, 3D active spheres and 2D/3D active rods of aspect ratio $l/\sigma=2$. A stability analysis of the homogeneous isotropic phase allows us to predict the onset of MIPS based on the effective swimming speed and rotational diffusion of the particles. Both methods find suppression of MIPS as the particle shape is elongated. We propose a suppression mechanism based on the duration of collisions, and argue that this mechanism can explain both the suppression of MIPS found here for rodlike particles and the enhancement of MIPS found for particles with Vicsek interactions.
\end{abstract}


\maketitle 

\section{Introduction}
\label{sec:Introduction}
Recently, there has been an increased interest in the thermodynamics of what has been coined active matter: systems formed by agents that can convert ambient or external energy into kinetic energy. These systems are diverse. Many are biological in nature: birds, fish, cells and bacteria all convert some form of ambient chemical energy into kinetic energy. Synthetic systems also exist in the form of colloidal particles that self-propel, typically by diffusiophoresis \cite{Theurkauff2012,Palacci2013,Buttinoni2013,Ginot2018}. All these active-matter systems are driven so far out of equilibrium that current theories of statistical thermodynamics fail to describe many of their properties. The dynamics of active-matter systems can be modeled quite easily. However, it has unfortunately proven to be very difficult to relate the dynamics to their steady states or to any kind of probability distribution. If we could apply the concepts of equilibrium statistical thermodynamics to active systems, it would greatly improve our ability to describe and predict the structure and behaviour of these systems. Some progress has been made in this regard. For instance, for self-propelled disks, the pressure has been defined \cite{Solon2015,Patch2017}, the glass transition has been investigated \cite{Fily2014a}, and equations of state have been constructed \cite{Levis2017}. Furthermore, the work of Cates and Tailleur contributes significantly in constructing effective free energies for these nonequilibrium systems \cite{Cates2015}.\\
For active 2D disks or 3D spheres, two well-studied model systems, there is often a parameter regime in which the system demixes into a dense and a dilute region. This phase separation closely resembles the well-known gas-liquid coexistence found in, for instance, water or Lennard-Jones systems. Unlike the gas-liquid phase separation, however, the clustering in active matter occurs because of the motility rather than the particle-particle attractions. Hence, the phenomenon has been coined motility-induced phase separation (MIPS). Recently, MIPS has been studied extensively: it has been identified for both active Brownian \cite{Fily2012,Redner2013,Fily2014,Marchetti2016} and run-and-tumble particles \cite{Tailleur2008}, its nucleation has been studied \cite{Redner2016}, its interface has been shown to allow for a negative surface tension \cite{Bialke2015a}, it has been derived from equations of state \cite{Patch2017,Levis2017}, and from nonequilibrium thermodynamics theories \cite{Cates2015}.\\
All of the above studies logically constrained themselves to the simplest possible model systems, in which particles interact either through hard-particle excluded-volume interactions or through short-range repulsions. Importantly, such models contain no torques. Studies that do include torques typically fall into two categories. The first uses particles with Vicsek-like alignment interactions \cite{Barre2014,Sese-Sansa2018}, which mimic a visual alignment mechanism, such as for birds or fish. The second uses particles with an anisotropic, typically rodlike shape \cite{Peruani2006,Ginelli2010,Peruani2015,Weitz2015,Velasco2018}. This most closely mimics bacteria, whose alignment arises simply from bumping into one another. While studies of active rods reveal a zoo of nonequilibrium phases, they do not exhibit MIPS; there seems to be no parameter regime for which there is a separation into dense and dilute regions without strong alignment. Naturally, this raises questions such as: why does MIPS occur for 2D disks and 3D spheres, but not for 2D and 3D rods? How anisotropic or rod-like must a particle be for MIPS to disappear? In this paper, we will address these questions by both simulations and theory.\\ 
To address these questions numerically, we need a model system which exhibits MIPS, and a means to identify MIPS when it occurs. Section \ref{sec:Methods} describes both the Active Brownian Particle model we use, and the modified cluster algorithm we apply to identify MIPS. In Section \ref{sec:Theory} we present an analytical criterion for the onset of MIPS, on the basis of a stability analysis of density fluctuations in the homogeneous isotropic phase, with the full derivation presented in the Appendix. In Section \ref{subsec:PhaseDiagrams} we discuss the phase diagrams for the 2D disks, 3D spheres and 2D and 3D rods, showing unambiguously that MIPS indeed disappears for increasing aspect ratio. Subsequently, we discuss the mechanism behind this suppression in Section \ref{subsec:MIPSsuppression}. Section \ref{sec:Conclusions} then concludes this paper by discussing the influence of torque on MIPS in a more general context.

\section{Computational methods}
\label{sec:Methods}

\subsection{Active Brownian Particles}
\label{subsec:ABP}
\begin{figure}[h]
\includegraphics[width=0.75\columnwidth]{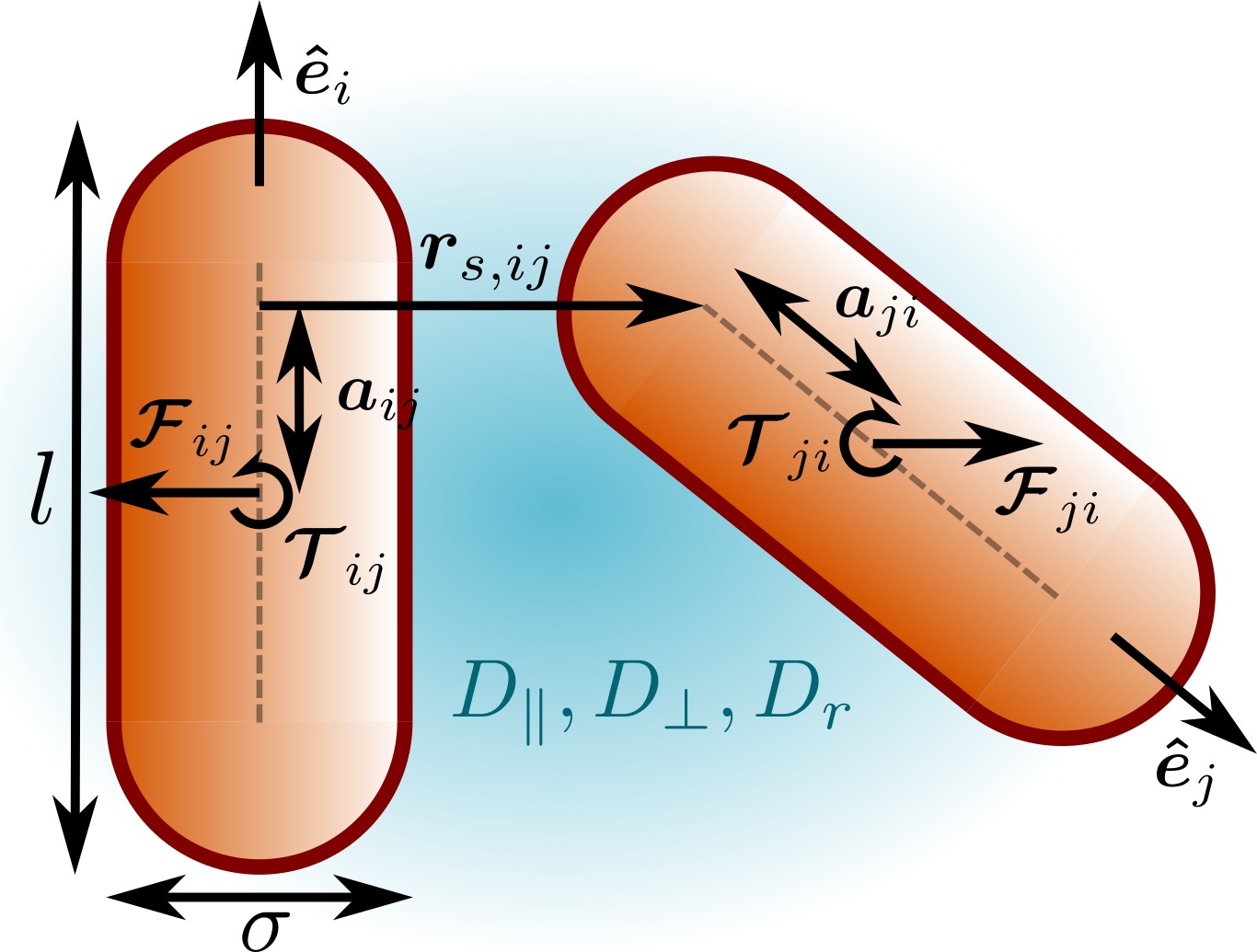}
\caption{\label{fig:BDSchematic} Schematic representation of the model. Particles are 2D or 3D spherocylinders of diameter $\sigma$ and head-to-tail length $l$, self-propelled with a velocity $v_0$ in their forward direction $\bm{\hat{e}}$. They interact based on their core-to-core distance $\bm{r_{s,ij}}$, causing repulsive forces $\bm{\mathcal{F}}_{ij}$ and torques $\bm{\mathcal{T}}_{ij}$. Additionally, they diffuse rotationally with diffusion constant $D_r$, and translationally along their long and short axis with diffusion constants $D_{\|}$ and $D_{\bot}$, respectively.}
\end{figure}

Using Brownian Dynamics (BD) simulations, we study a system of $N$ spherocylinder-shaped active Brownian particles (ABP) of head-to-tail length $l$ and diameter $\sigma\leq l$ in a periodic area $A$, self-propelling with a velocity $v_0$ along their long axis $\bm{\hat{e}}$. The particles are subject to rotational and translational noise, with rotational diffusion constant $D_r$ and translational diffusion tensor $\bm{D}_t=D_{\|}\bm{\hat{e}}\bm{\hat{e}} + D_{\bot}(\bm{\mathcal{I}}-\bm{\hat{e}}\bm{\hat{e}})$, with parallel and perpendicular components $D_\|$ and $D_\bot$, respectively. For such a 2D system, shown schematically in Fig. \ref{fig:BDSchematic}, the overdamped Langevin equations are given by
\begin{align}
\partial_t \bm{r}_i &=
   v_0\bm{\hat{e}}_i +
   \beta \bm{D}_{t,i} \cdot \sum_{i\neq j} \bm{\mathcal{F}}_{ij} +
   \sqrt{2 \bm{D}_{t,i}} \cdot \bm{\Lambda}^t_i; \\
\partial_t \theta_i &=
   \beta D_r \sum_{i\neq j} \mathcal{T}_{ij} +
   \sqrt{2 D_r} \Lambda^r_i,
\end{align}
where $i=1,...,N$ is the particle label, $\bm{r}_i$ is the position of particle $i$, $\bm{\hat{e}}_i=(\cos\theta_i, \sin\theta_i)$ the particle orientation, and $\beta=1/k_B T$. The force $\bm{\mathcal{F}}_{ij}$ and torque $\mathcal{T}_{ij}$ are due to particle-particle interactions. We assume fluctuation-dissipation to hold on the scale of individual particles, such that the translational and rotational noise terms $\Lambda^{t,\alpha}_i$ and $\Lambda^r_i$, respectively, are Gaussian distributed random numbers with zero mean and unit variance, i.e.:
\begin{align}
\langle \Lambda_i \rangle &= 0 ;\\
\langle \Lambda_i^\alpha(t) \Lambda_j^\beta(t') \rangle &= \delta_{ij}\delta_{\alpha\beta}\delta(t-t').
\end{align}
To describe excluded-volume interaction between particles $i$ and $j$, we let the forces $\bm{\mathcal{F}}_{ij}=(\partial u_{WCA}(r_{s,ij})/\partial r_{s,ij}) \bm{\hat{r}}_{s,ij}$ be the result of a short-range pairwise repulsive WCA potential $u_{WCA}(r_{s,ij})$ acting on the shortest distance $r_{s,ij}$ between particle cores:
\begin{align}
& u_{WCA}(r_{s,ij})  = \\
&\begin{cases}
\ 4\epsilon\left[ \left(\frac{\sigma}{r_{s,ij}}\right)^{12}-\left(\frac{\sigma}{r_{s,ij}}\right)^6\right]+\epsilon   &   \mathrm{if}\ \   r_{s,ij}<2^{1/6}\sigma;\\
\ 0   &   \mathrm{if}\ \   r_{s,ij}\geq2^{1/6}\sigma \nonumber.
\end{cases}
\end{align}
For disks ($l/\sigma=1$), the distance $r_{s,ij}$ is simply the distance between their centers. For $l/\sigma>1$ the cores of the particles are no longer points, but lines. The distance $r_{s,ij}$ then corresponds to the shortest distance between these two line segments. The torques $\mathcal{T}$ are calculated from the forces by $\bm{\mathcal{T}}_{ij}=\bm{a}_{ij}\times\bm{\mathcal{F}}_{ij}$, where $\bm{a}_{ij}$ is the lever arm for the applied force $\bm{\mathcal{F}}_{ij}$ on rod $i$ by rod $j$. For each pair of particles, both the shortest distance $r_{s,ij}$ and the lever arms $\bm{a}_{ij}$ are calculated using the algorithm described in Ref. \cite{Vega1994}. In 2D this torque always points out of plane, so we only need to consider its scalar magnitude $\mathcal{T}$ in the equations of motion.\\

This 2D model easily generalizes to 3D: aside from vectorial quantities now being three- rather than two-dimensional, we must now also consider the direction of the torque. For convenience, we also switch to vector notation in the orientational equation of motion. The equations of motion in 3D are thus:
\begin{align}
\partial_t \bm{r}_i &=
   v_0\bm{\hat{e}}_i +
   \beta \bm{D}_{t,i} \cdot \sum_{i\neq j} \bm{\mathcal{F}}_{ij} +
   \sqrt{2 \bm{D}_{t,i}} \cdot \bm{\Lambda}^t_i; \\
\partial_t \bm{\hat{e}}_i & =
   \beta D_r \sum_{i\neq j} \bm{\mathcal{T}}_{ij} \times \bm{\hat{e}}_i  +
    \sqrt{2 D_r} (\bm{\hat{e}}_i \times \bm{\Lambda}^r_i).
\end{align}
We nondimensionalize the 2D and 3D system by expressing all distances in units of the particle diameter $\sigma$, all energies in terms of the thermal energy $k_B T$, and all units of time in terms of $\tau=1/D_r$.\\

\subsection{Choice of model parameters and additional assumptions}
\label{subsec:Parameters}
For our investigation, we will study the influence of four parameters: the dimensionality $d=2$ and $d=3$, the aspect ratio $p=l/\sigma$, the packing fraction $\phi=N((\pi/4) \sigma^2 + (l-\sigma)\sigma) /A$ ($\phi=N((\pi/6) \sigma^3+(\pi/4)(l-\sigma)\sigma^2)/V$ in 3D) and the P\'eclet number $\text{Pe}=v_0/\sigma D_r$. Note that literature sometimes defines the P\'eclet number in terms of the translational diffusion instead. The diffusion constants $D_{\|}$ and $D_{\bot}$ for rodlike particles can be calculated from simulations including hydrodynamics as in e.g. Ref. \cite{Bet2017} or, for short spherocylinders, approximated by the exact results for ellipsoids (\cite{Perrin1934}, see also SI). We found that the influence of this change of diffusion constants is negligible for the aspect ratio range we look at, so for simplicity we will set $D_{\|}=D_{\bot}=D_t=\sigma^2 D_r/3$ from now on. This choice corresponds to the correct ratio between translational and rotational diffusion for 3D spheres.\\
Some care is required in the way we vary the P\'eclet number. The most straightforward way is to simply vary it by changing the self-propulsion velocity $v_0$. However, if we do this and keep the pair interaction strength fixed, the ratio between active and interaction forces will depend on the P\'eclet number. The result of changing this ratio is that the particle interaction effectively becomes softer as the P\'eclet number increases. In the extreme case, MIPS may even disappear for high enough P\'eclet numbers. Earlier work has remarked on this subtlety of varying the P\'eclet number \cite{Stenhammar2014,Bialke2013}. As our aim is not to provide quantitative but only qualitative data on the phase behaviour, we nevertheless use the straightforward approach by fixing $\epsilon=24k_B T$ and changing the P\'eclet number by varying $v_0$.\\

\subsection{Identifying motility-induced phase separation by clustering regions of similar density}
\label{subsec:IdentifyingMIPS}
\begin{figure}[h]
\includegraphics[width=.98\columnwidth]{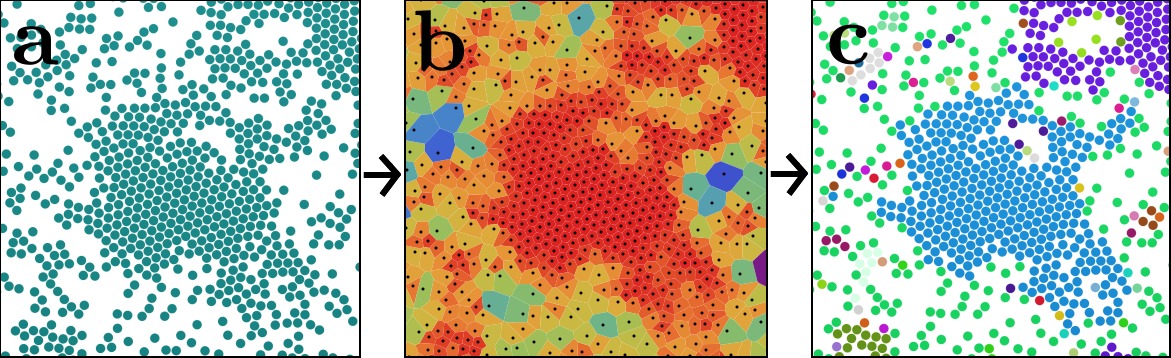}
\caption{\label{fig:Cluster} Representation of our clustering algorithm. From unlabeled  coordinates (a), construct a Voronoi tesselation and obtain local densities (b), then use these to create clusters of particles with similar density (c).}
\end{figure}

MIPS is a separation of a system of self-propelled particles into a dense and a dilute region. While it can be identified quite readily from visual inspection of particle configurations, it is also useful to have a more quantitative method. Two of these methods are common. The first is to measure the distribution of the local density: for a homogeneous system, such a distribution is unimodal, while for a phase-separated system it is bimodal \cite{Suma2014,Speck2016a,Bruss2018}. However, such a distribution can not tell us whether the system has separated into one or into multiple domains, which means it cannot distinguish between micro- and macrophase separation. This distinction becomes important for rods.\\
The other method is to group particles together into clusters based on a distance cutoff and to determine the fraction $f_\text{cl}$ of particles in the largest cluster \cite{Fily2012,Buttinoni2013,Prymidis2016}. Since MIPS eventually forms one large, dense cluster in a very dilute background gas, $f_\text{cl}\rightarrow 1$ for MIPS for large enough systems, while for a homogeneous fluid $f_\text{cl}\rightarrow 0$. This latter method requires a cutoff distance that specifies whether particles are close enough to belong to the same cluster. In practice, we found that there is no single cutoff distance that yielded reasonable results for the resulting cluster fraction across all shapes and densities we wish to study.\\
To solve this problem, we developed a slightly different clustering method, shown schematically in Fig. \ref{fig:Cluster}. From the particle positions (Fig. \ref{fig:Cluster}a), we make a Voronoi construction. This provides us not only with a parameter-free way to define neighbouring particles, but also with a means of measuring the local packing fraction: $\phi_l = ((\pi/4) \sigma^2 + (l-\sigma)\sigma)/A_v$ or $((\pi/6) \sigma^3+(\pi/4)(l-\sigma)\sigma^2)/V_v$, with $A_v$ and $V_v$ the area (2D) or volume (3D) of the Voronoi cell (colors in Fig. \ref{fig:Cluster}b). Our requirements for two particles to belong to the same cluster are then that (a) their Voronoi cells are connected and (b) they both have a local packing fraction that is either lower or higher than the mean packing fraction $\phi$ by a certain cutoff $\Delta \phi$. Using this method, we create clusters of similar local density (Fig. \ref{fig:Cluster}c). We choose $\Delta\phi=0.025$, as we found through trial and error that this cutoff allows us to meaningfully distinguish between homogeneous states with $f_\text{cl}<0.5$ and phase-separated states with $f_\text{cl}\geq0.5$ for all aspect ratios and P\'eclet numbers of interest and for nearly all densities, both in $d=2$ and $d=3$. Note that $f_\text{cl}$ is not guaranteed to go to zero in the homogeneous phase when using this definition of clusters due to density fluctuations, but $f_\text{cl}=0.5$ still offers a reasonable threshold.\\

\section{An analytical criterion for the onset of MIPS}
\label{sec:Theory}
Having described the means to obtain and identify MIPS numerically, we now describe an analytical criterion for the onset of MIPS. We are aware of three ways to obtain such a criterion: by considering the particle flux balance between a dense cluster and a dilute gas phase \cite{Redner2013,Redner2013a}, by constructing an effective free energy and proceeding as in equilibrium \cite{Cates2013,Cates2015}, and by a stability analysis of density fluctuations of the homogeneous isotropic phase \cite{Bialke2013,Speck2015}. All three methods have previously been used for torque-free systems. We extend the mean-field-like third method laid out in \cite{Speck2015} to 3D systems with torque. The derivation of this extension is given in the Appendix. In short, we map our system to an active ideal gas, where the effect of the many-body forces and torques is subsumed into a modified, \textit{effective} swim speed $v^\text{eff}$, rotational diffusion $D_r^\text{eff}$, and translation diffusion $D_t^\text{eff}$. These effective constants then depend on the mean density $\bar{\rho}$ and input swimming speed $v_0$. By doing this mapping, we effectively make two approximations: the only effect of the interparticle forces $\bm{\mathcal{F}}$ is to slow particles down, and the only effect of the torques $\bm{\mathcal{T}}$ is to change the rate at which particles change their orientation. The former is a good approximation in the absence of structural order, the latter in the absence of orientational order. Both approximations become poorer at higher densities, where structure and alignment become important.\\
As will be derived in Appendix 2, the evolution of long-range density perturbations $\delta\rho(\bm{r},t)$ for this active ideal gas is given by 
\begin{equation}
\partial_t \delta \rho (\bm{r},t) \approx \mathcal{D}_{\delta\rho}(\bar{\rho},v_0)\bm{\nabla}^2\delta\rho(\bm{r},t).
\label{eq:DensityDiffusion}
\end{equation}
This is a diffusion equation with a diffusion coefficient $\mathcal{D}_{\delta\rho}$ given by:
\begin{equation}
\mathcal{D}_{\delta\rho}(\bar{\rho},v_0) = D_t^\text{eff} + \frac{v^\text{eff}(2v^\text{eff}-v_0)}{d(d-1)D_r^\text{eff}},
\label{eq:DensityDiffusionConstant}
\end{equation}
where $d$ now indicates the dimensionality. In agreement with literature and as detailed in the SI, we confirm from our simulations that $D_t^\text{eff}\approx D_t$ for spheres to a reasonable approximation, similar to what was reported in Ref. \cite{Speck2015} for disks. This result extends to 3D rods as well. Thus, we set $D_t^\text{eff}=D_t$ from here on.\\
The effective constants $v^\text{eff}$ and $D_r^\text{eff}$ can now be found in two ways: we can either formulate closed-form equations for these effective constants, or we could measure them in some way. We choose the latter method, and determine their value from the following correlation functions:
\begin{align}
\label{eq:veff} \langle \bm{\dot{r}}_i (t)\cdot \bm{\hat{e}}_i(t) \rangle & = v^{\text{eff}}; \\
\label{eq:Dreff} \langle \bm{\hat{e}}_i(t)\cdot\bm{\hat{e}}_i(0) \rangle & = \exp(-(d-1)D_r^{\text{eff}}t),
\end{align}
which measure the effective velocity in the direction of self-propulsion and how quickly a particle loses its orientation, respectively. In other words, we can measure $v^\text{eff}$ and $D_r^\text{eff}$ by simulating a (small) system in the homogeneous isotropic phase.

\section{Results \& Discussion}
\label{sec:Results}

\begin{figure*}[t]
\includegraphics[width=.98\columnwidth]{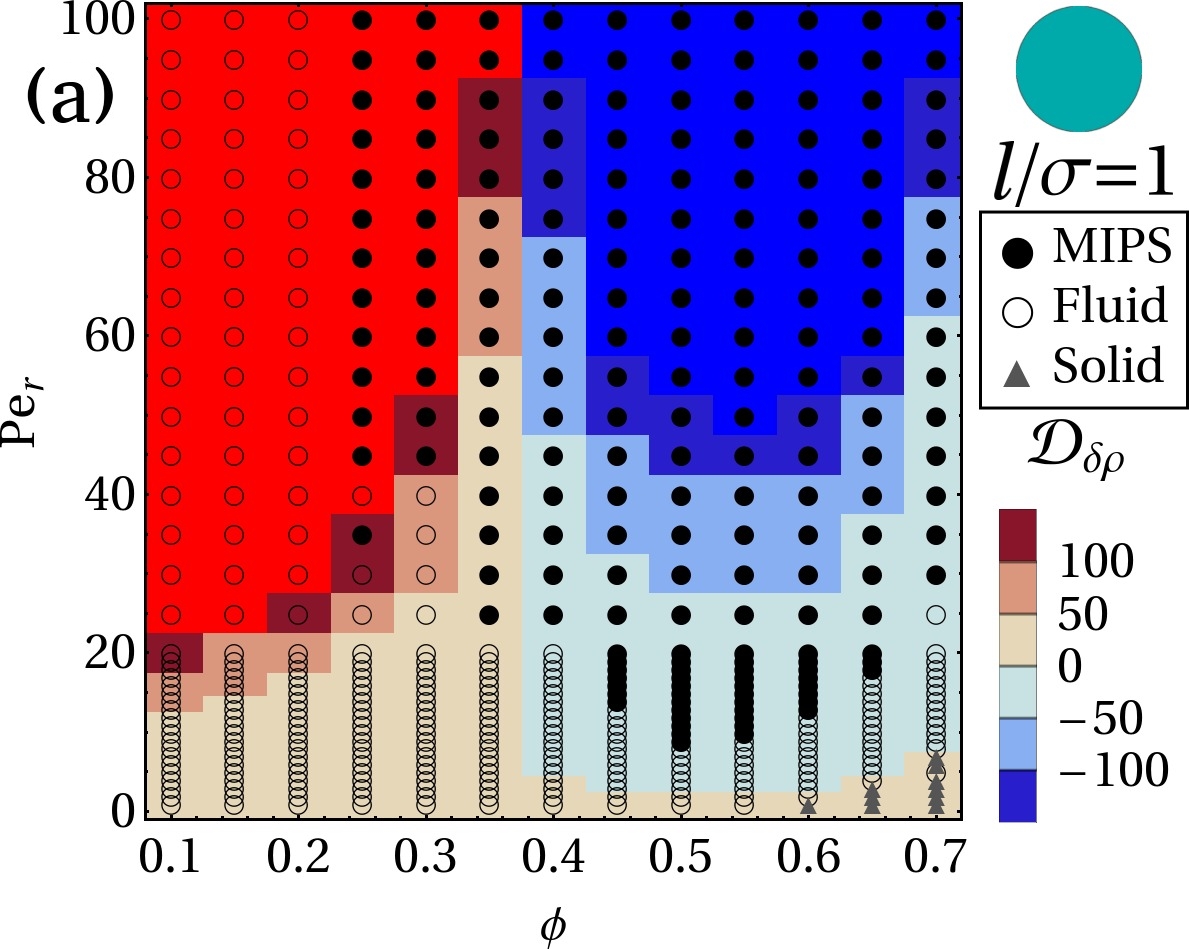}
\hspace{0.04\textwidth}
\includegraphics[width=.98\columnwidth]{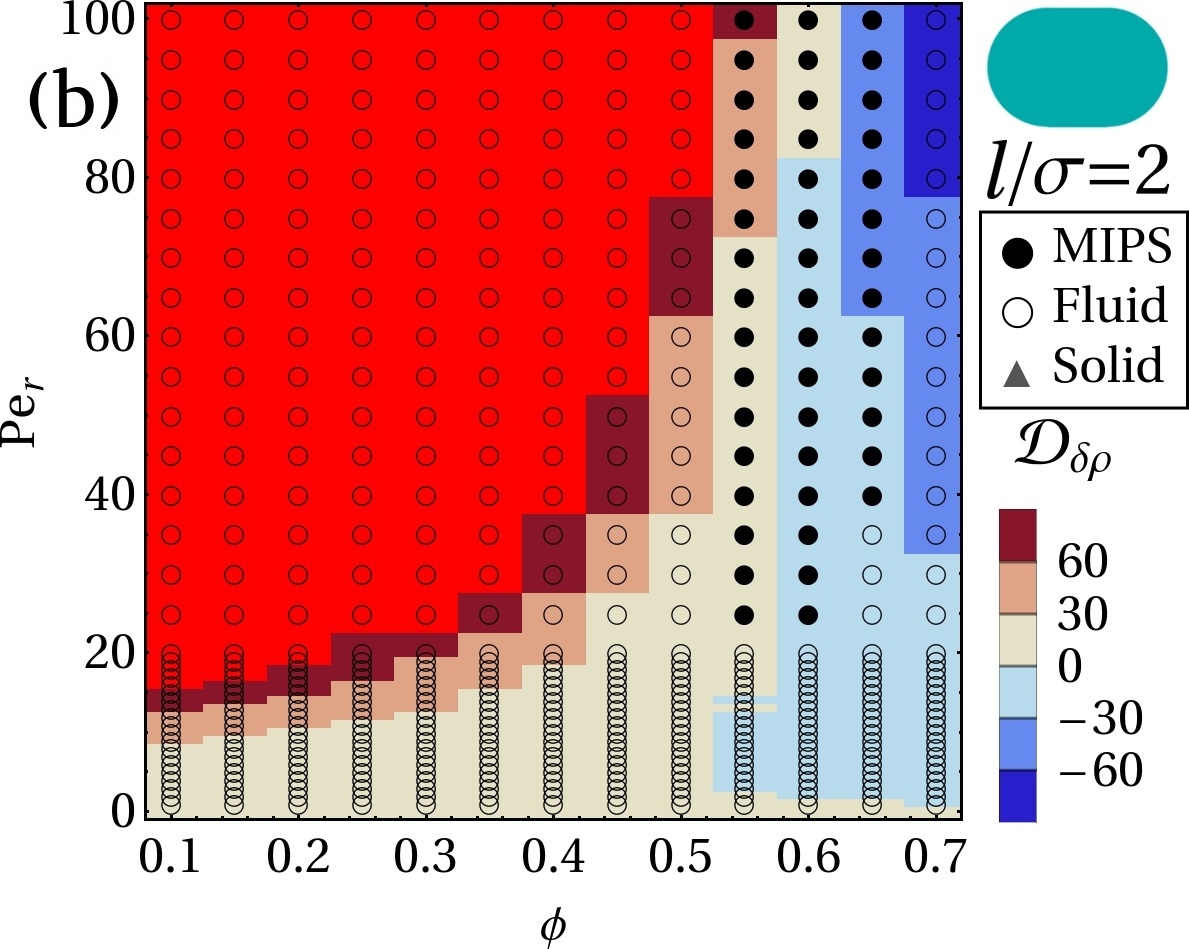} 
\caption{\label{fig:2DPhaseDiagram} Phase diagram of 2D self-propelled disks (a) and rods of aspect ratio $l/\sigma=2.0$ (b), for different P\'eclet numbers $\text{Pe}$ and packing fractions $\phi$. Data points indicate the resulting phase of $N=10^4$ particles as obtained from Brownian dynamics simulations, in which we distinguished MIPS, fluid and solid phases. The colors indicate the diffusion constant of density fluctuations $\mathcal{D}_{\delta\rho}$. Spinodal decomposition to a MIPS state is then predicted where $\mathcal{D}_{\delta\rho}<0$.}
\end{figure*}

To explore the MIPS-related phase behaviour, we performed Brownian Dynamics simulations with $N=10^4$ particles in the packing-fraction range $0.1\leq\phi\leq0.7$, and the P\'eclet-number range $1\leq\text{Pe}\leq100$ (2D)  and $1\leq\text{Pe}\leq150$ (3D). This spans the entire density range from the fluid regime to just below the hexatic/solid regimes \cite{Bates2000,Digregorio2018}. The P\'eclet range spans from below the MIPS critical point to high enough P\'eclet that the MIPS region attains a near-constant width in density \cite{Speck2014,Stenhammar2014}. We also performed smaller simulations to measure the effective constants $v^\text{eff}$ and $D_r^\text{eff}$. The initial state for all simulations was one with random positions and orientations. Using only a limited number of particles ($N=100$) ensures that the system remains in the homogeneous isotropic phase regardless of density or activity. Of course, these smaller simulations  suffer from finite-size effects. The supplementary information contains a more detailed analysis of these finite-size effects.\\
We now turn to answer the questions posed in the introduction. How elongated do particles have to be to not display MIPS? And what is the mechanism that suppresses MIPS for rods? \\
 
\subsection{Phase diagrams of self-propelled disks, spheres and rods}
\label{subsec:PhaseDiagrams}
Before we can appreciate how the phase behaviour changes with aspect ratio, we must first establish the relevant features of MIPS for isotropic particles. Let us start in 2D.\\

While the phase boundaries of MIPS for disks have been studied by a number of authors \cite{Redner2013,Speck2014,Solon2015,Speck2015,Levis2017}, a comprehensive study that also includes the high-density hexatic and solid phases has only appeared quite recently \cite{Digregorio2018}. In this study, the authors report not only the commonly-reported U-shaped MIPS region (in the density-activity plane), but also that there a narrow density regime wherein a hexatic phase can be found. This regime spans from the passive system ($\text{Pe}=0$) to connect to the MIPS region ($\text{Pe}\sim100$). Note that Ref. \cite{Digregorio2018} defines the P\'eclet number in terms of the active force, while we express it in terms of $v_0$ and $D_r$---this shifts the scale by a constant factor of $D_r\sigma^2/D_t=3$. We can expect to find qualitatively similar features here. Quantitatively, the phase boundaries will be shifted somewhat because of differences in the repulsive pair potential: Ref. \cite{Digregorio2018} uses $U\propto (\sigma/r)^{64}-(\sigma/r)^{32}$, while we use $U\propto (\sigma/r)^{12}-(\sigma/r)^{6}$. Our softer potential decreases the size of the liquid-hexatic coexistence region \cite{Kapfer2015}. Given the similar temperatures ($k_B T/\epsilon = 1/24$, versus $k_B T/\epsilon = 1/20$ in Ref. \cite{Digregorio2018}), the slightly longer range of the WCA potential (cutoff radius $r/\sigma=2^{1/6}\sim1.12$, as opposed to $r/\sigma=2^{1/32}\sim1.02$ in Ref. \cite{Digregorio2018}) will shift the solid phases to slightly lower packing fractions. However, an in-depth comparison of the high-density phase boundaries falls outside the scope of this paper. Instead, we will merely identify the solid-like phases by looking at where the effective velocity $v^\text{eff}$ becomes vanishingly small ($v^\text{eff}<0.1\sigma D_r$). Although this is not a very accurate measure, but it serves to crudely distinguish the solid or hexatic phase from the fluid and MIPS phases, at least at low self-propulsion. We use this criterion for all phase diagrams throughout this paper. With this information in mind, let us now consider the phase diagrams in Figure \ref{fig:2DPhaseDiagram}.\\

Figure \ref{fig:2DPhaseDiagram} shows phase diagrams in the P\'eclet number $\text{Pe}$ - packing fraction $\phi$ representation for 2D disks and rods that show both the MIPS region predicted on the basis of the stability analysis (blue-tinted region, $\mathcal{D}<0$), and the MIPS region found in the simulations using $N=10^4$ (black points). Both methods seem to indicate MIPS in roughly the same region, but there are a few notable differences. On the low density side, we also find MIPS outside of the predicted spinodal region. Making the analogy with the gas/liquid phase separation, we would expect MIPS in this region to then occur through nucleation and growth. Is this also the case?\\

A simple way to see if MIPS forms through a nucleation process is to look at domain growth, which we can track using a time series of cluster fraction $f_{\text{cl}}$ defined in Section \ref{subsec:IdentifyingMIPS}. If the system immediately decays from an isotropic to a MIPS state, this fraction will likewise increase immediately. If, on the other hand, the system stays in the fluid state for a prolonged period of time, only to later transition into MIPS through a nucleation process, $f_{\text{cl}}$ will retain the value corresponding to the fluid for a finite time.\\

\begin{figure}[h]
\includegraphics[width=.99\columnwidth]{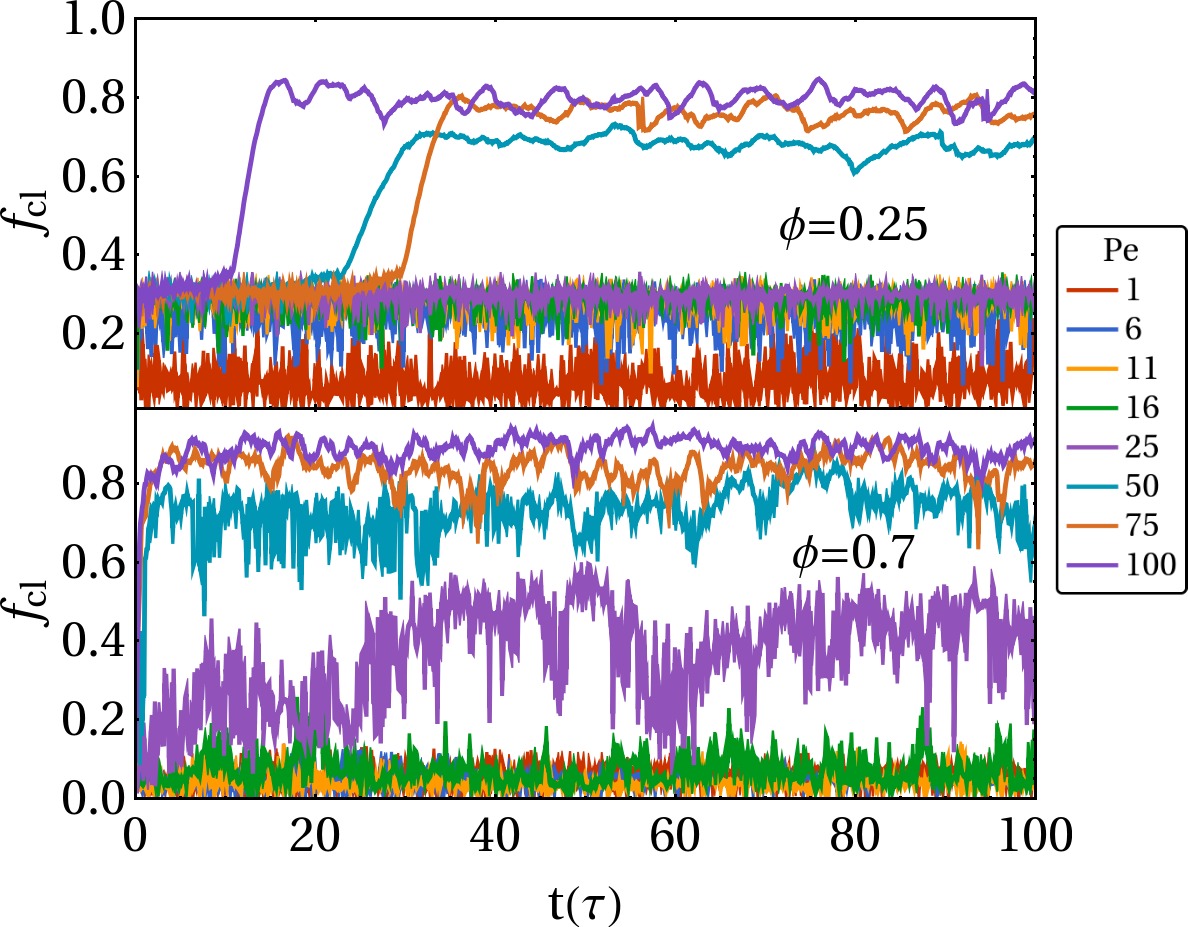}
\caption{\label{fig:Nucleation} Time series of the largest cluster fraction $f_{\text{cl}}$ for active disks ($l/\sigma=1$). At low density $\phi=0.25$, the system occasionally only clusters after a significant amount of time ($t>10\tau$), suggesting that the transition is triggered by a rare nucleation event. At high density $\phi=0.7$, this is never the case\textemdash only spinodal decomposition is observed.}
\end{figure}

Figure \ref{fig:Nucleation} compares the time evolution of the fraction $f_\text{cl}$ of particles in the largest cluster for a number of P\'eclet numbers at two different densities: one on the low density side of the MIPS regime at $\phi=0.25$ and one on the high density side at $\phi=0.7$. On the low density side and outside of the predicted spinodal region, the cluster fraction can stay constant for a significant amount of time ($t>30\tau$) before transitioning to a MIPS state. On the high density side of the MIPS region, such a delay is absent. The stability analysis predicts spinodal decomposition in this regime, and the cluster growth agrees. This asymmetry is consistent with the findings of Speck et al. \cite{Speck2015}, who report that the MIPS transition is discontinuous at low densities, but continuous at high densities. \\

There is also a discrepancy between the stability analysis and the large-scale simulation at low P\'eclet numbers. This is to be expected:  in this region the fluid-MIPS transition is continuous, and the difference in density between the coexisting phases is small when we close to a critical point. Consequently, distinguishing between clusters of particles is difficult, and the exact choice of cluster fraction threshold $f_\text{cl}$ can shift the boundary quite a bit in this region.\\

\begin{figure}
\includegraphics[width=\columnwidth]{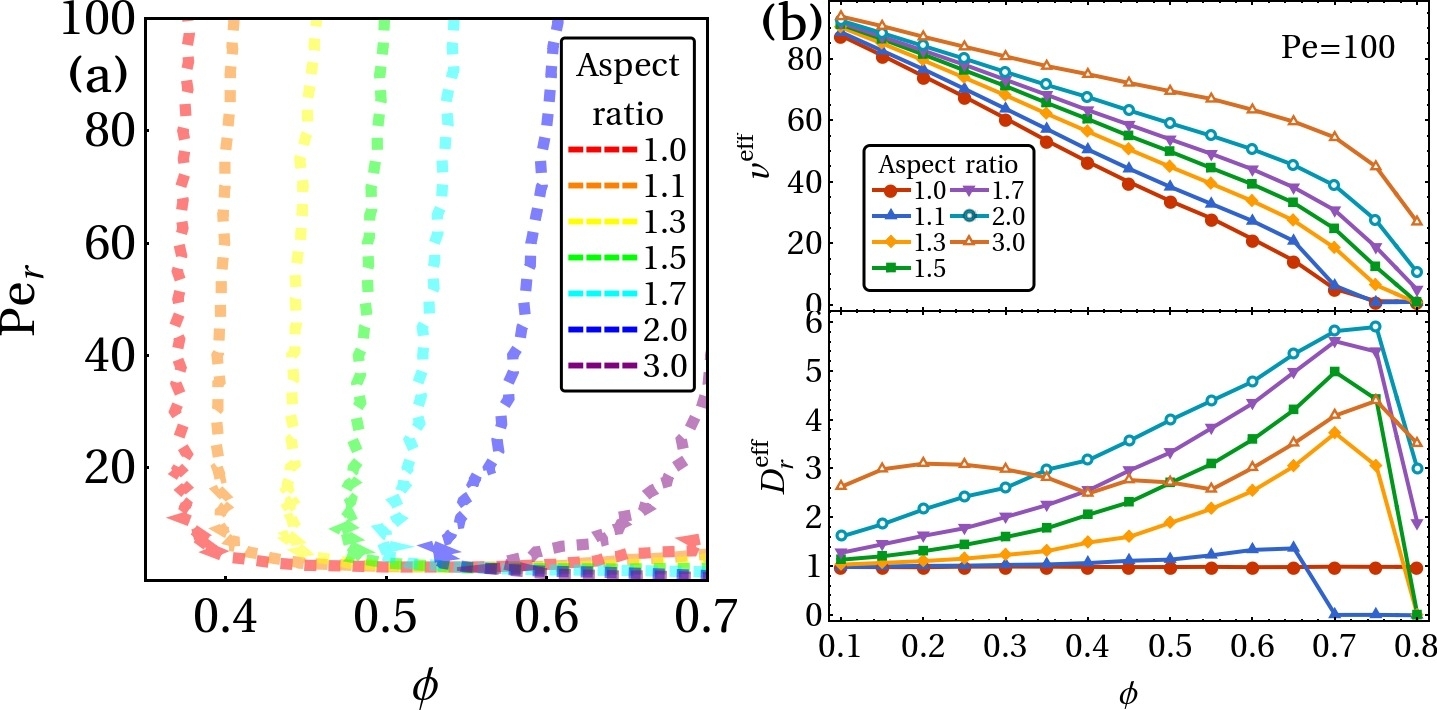}
\caption{\label{fig:2DAspectRatioDiagram} Spinodal lines for 2D active rods as predicted from Eq. \ref{eq:DensityDiffusionConstant}, as a function of aspect ratio (a), and the corresponding effective self-propulsion velocity $v^\text{eff}$ and rotational diffusion $D_r^\text{eff}$ at P\'eclet number $\text{Pe}=100$ as a function of packing fraction (b). At high activity, the effective self-propulsion decreases more slowly with density, while the rotational diffusion is enhanced.}
\end{figure}

Having identified the most important features of the phase diagram for active disks, let us now turn to rods and see how these features change. Figure \ref{fig:2DPhaseDiagram}b shows the phase diagram in the density-activity representation for rods with an aspect ratio of $l/\sigma=2$, using the same density and activity ranges as for the disks. The most obvious difference with the rods is that the MIPS region is now both shifted to higher densities and much narrower. The predictions of the stability analysis are worse for the rods: the predicted spinodal now lies in the middle of the simulated MIPS region. We find that the transition from fluid to MIPS now appears to be completely continuous---the system always starts clustering immediately, without any nucleation-like transient period. As can be seen from Figure \ref{fig:2DAspectRatioDiagram}a, the suppression is continuous with increasing aspect ratio, and it eventually pushes the fluid-MIPS transition into the regime where solid phases typically emerge.\\

\begin{figure*}[t]
\includegraphics[width=.99\columnwidth]{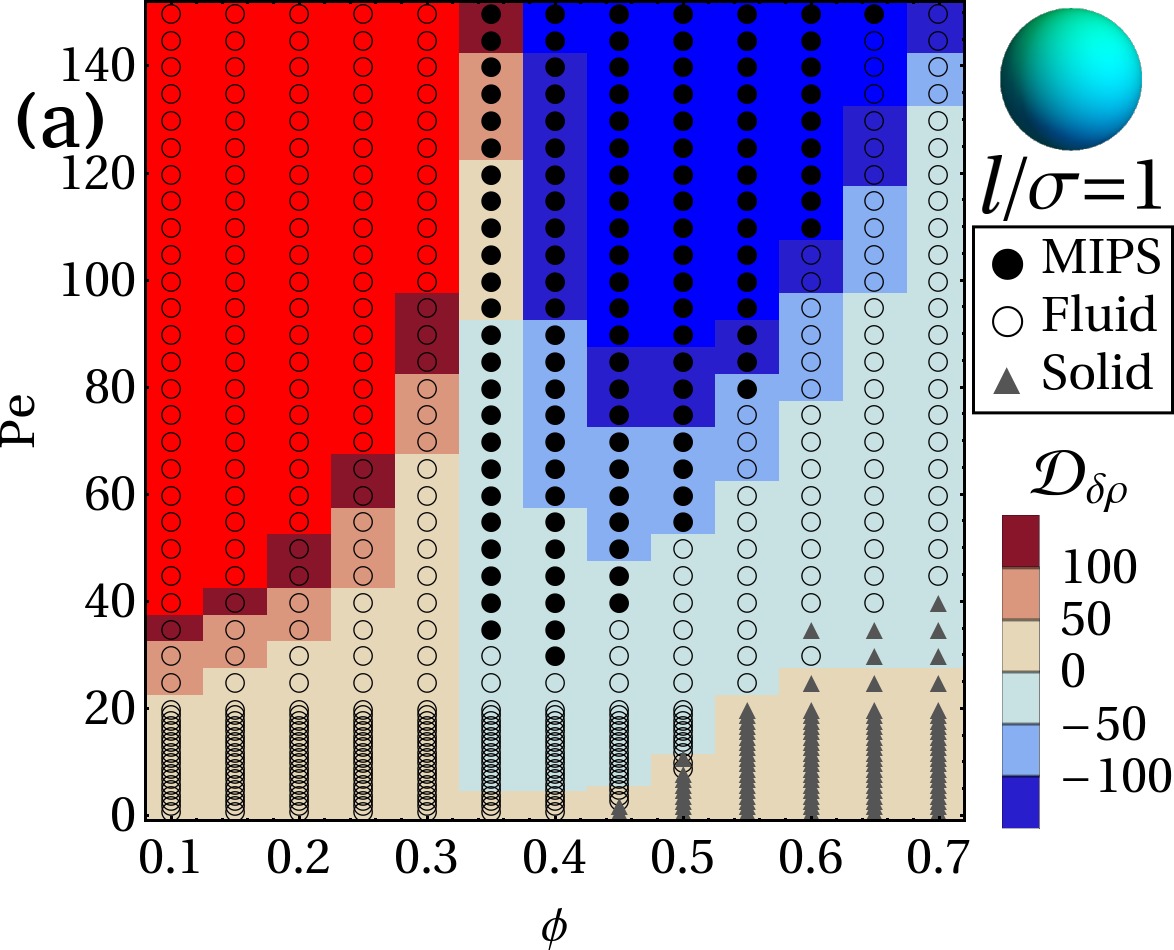} 
\includegraphics[width=.99\columnwidth]{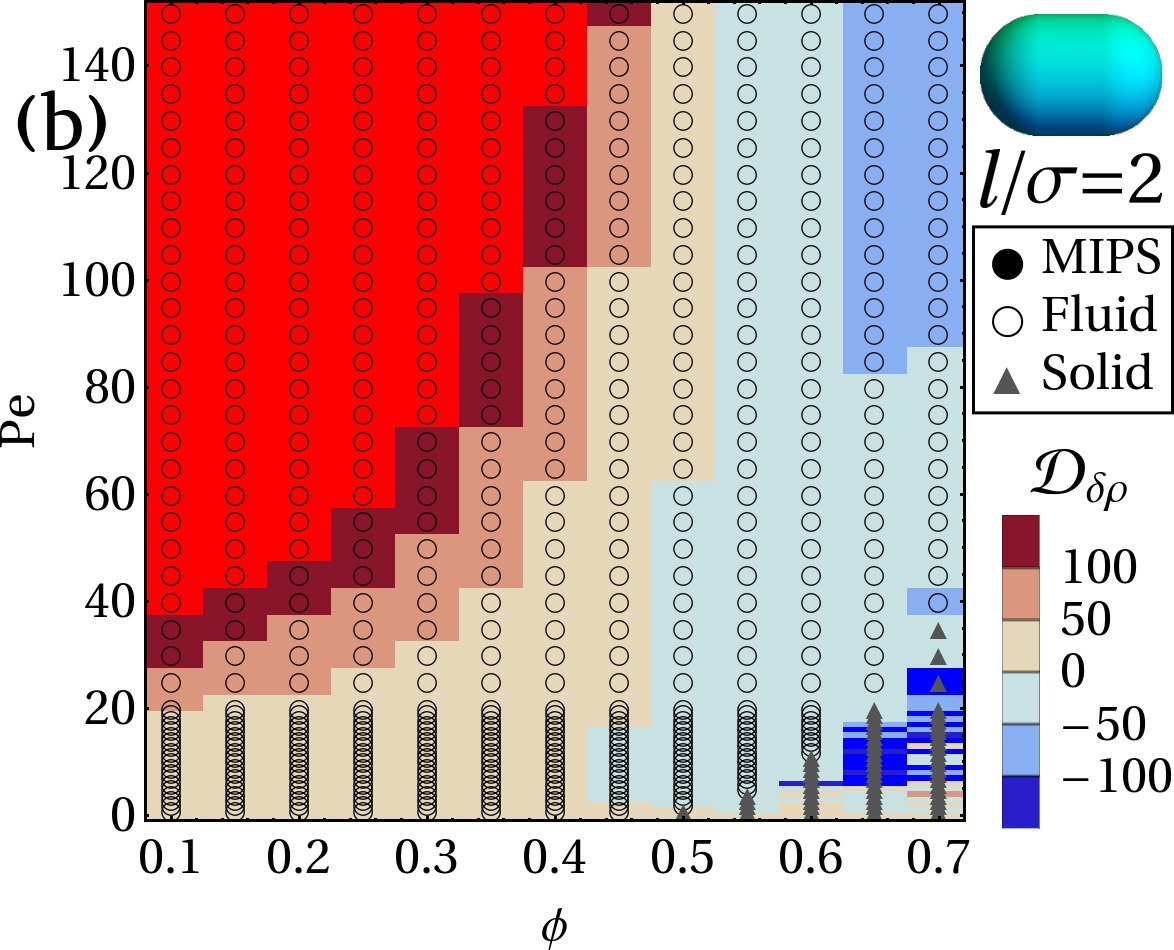} 
\caption{\label{fig:3DPhaseDiagram} Phase diagram of 3D self-propelled spheres (a) and rods of aspect ratio $l/\sigma=2.0$ (b), in the P\'eclet number $\text{Pe}$-packing fraction $\phi$ representation. Data points indicate the resulting phase of $N=10^4$ particles as obtained from Brownian dynamics simulations, in which we distinguished MIPS, fluid and solid phases. The colors indicate the diffusion constant of density fluctuations $\mathcal{D}_{\delta\rho}$. Spinodal decomposition to a MIPS state is then predicted to occur in the blue region where $\mathcal{D}_{\delta\rho}<0$. The small region of predicted instability in (b) under the points indicated as solid is an artefact of the fluid-solid transition there, where $D_{\delta\rho}$ fluctuates strongly as both $v^\text{eff}$ and $D_r^\text{eff}$ go to zero.}
\end{figure*}

Let us now see whether the 3D case is similar. Figure \ref{fig:3DPhaseDiagram} displays phase diagrams in the $(\phi,\text{Pe})$ representation, in Fig. \ref{fig:3DPhaseDiagram}a for 3D spheres and in Fig. \ref{fig:3DPhaseDiagram}b for 3D rods with $l/\sigma=2$. Somewhat unsurprisingly, they are similar to their 2D counterparts. The most important feature is retained: MIPS disappears when the aspect ratio is increased. The fluid gap we found in between the solid and MIPS phases is also present for the active spheres. However, there are notable differences between the 2D and 3D cases.\\

In contrast to the 2D case, we observe no region for the active spheres where the MIPS transition is discontinuous. All simulations that form MIPS appear to undergo immediate spinodal decomposition. This does not necessarily mean that there is no binodal region: it may simply be quite small or have low nucleation barriers. The density regime of the metastable region for 3D active spheres is not well understood. We are only aware of one comparable simulation study by Stenhammar et al. \cite{Stenhammar2014}, who looked at 2D and 3D active disks/spheres to study the influence of dimensionality. However, their binodal lines were defined as the density at which a high-P\'eclet system phase separated, which is not directly comparable to the metastable region we define here. Hence, further studies are needed to explain the difference in the width of the metastable region between $d=2$ and $d=3$.\\

Another difference is at high P\'eclet number, where the predicted MIPS region for the spheres continues to shift towards higher density, instead of moving towards a constant one. We believe this to be the behaviour that we discussed in Section 
\ref{subsec:Parameters}: for higher P\'eclet numbers the particles can approach each other closer due to the active forces, causing the effective diameter of the particles to decrease. This effect appears to be stronger in 3D than in 2D, presumably due to the increased coordination of each particle.\\

The final difference between the 2D and 3D cases is perhaps the most notable one: for the rods, MIPS has disappeared completely. Whatever mechanism suppresses MIPS, it appears to be stronger in 3D than in 2D. Curiously enough, the stability analysis still predicts MIPS in a significant portion of the phase diagram. This discrepancy, combined with its 2D counterpart, suggests that our theoretical approach breaks down for longer aspect ratios. We will see why this is the case in the next section, where we discuss the suppression mechanism.\\


Armed with the knowledge of these phase diagrams, can we now answer the first question posed in our introduction: ``How rodlike must a particle be for MIPS to disappear?" Only partially, unfortunately. Determining the exact aspect ratio where MIPS disappears turns out to be quite difficult. We now know that the nature of the suppression stems from the fluid-MIPS transition shifting to higher densities, but unfortunately our methods to identify MIPS are less reliable at higher densities. More importantly, however, when the particle interactions are not isotropic, MIPS is no longer defined unambiguously and multiple types of clustered phases are possible which all fit the present criteria. When we identify MIPS according to a) the system phase-separating into a single dense cluster in a background gas and b) this cluster having no net orientational order, there are still multiple realizations of such a system (see SI, figure S5), such as a dense cluster with large domains of oppositely oriented particles ($l/\sigma=1.1$) or a percolating cluster with low orientational order and many voids ($l/\sigma=1.3,\ 2.0$). Therefore, establishing the boundaries of MIPS at these higher densities requires a more careful consideration of both hexatic \cite{Digregorio2018} and orientational order \cite{Shi2018}. We leave this investigation to future work and instead, having established \textit{that} MIPS is suppressed when particles become elongated, we now turn to finding out \textit{why}.\\

\subsection{Torque-induced suppression of motility-induced phase separation}
\label{subsec:MIPSsuppression}

\begin{figure*}[t]
\includegraphics[height=.55\textheight]{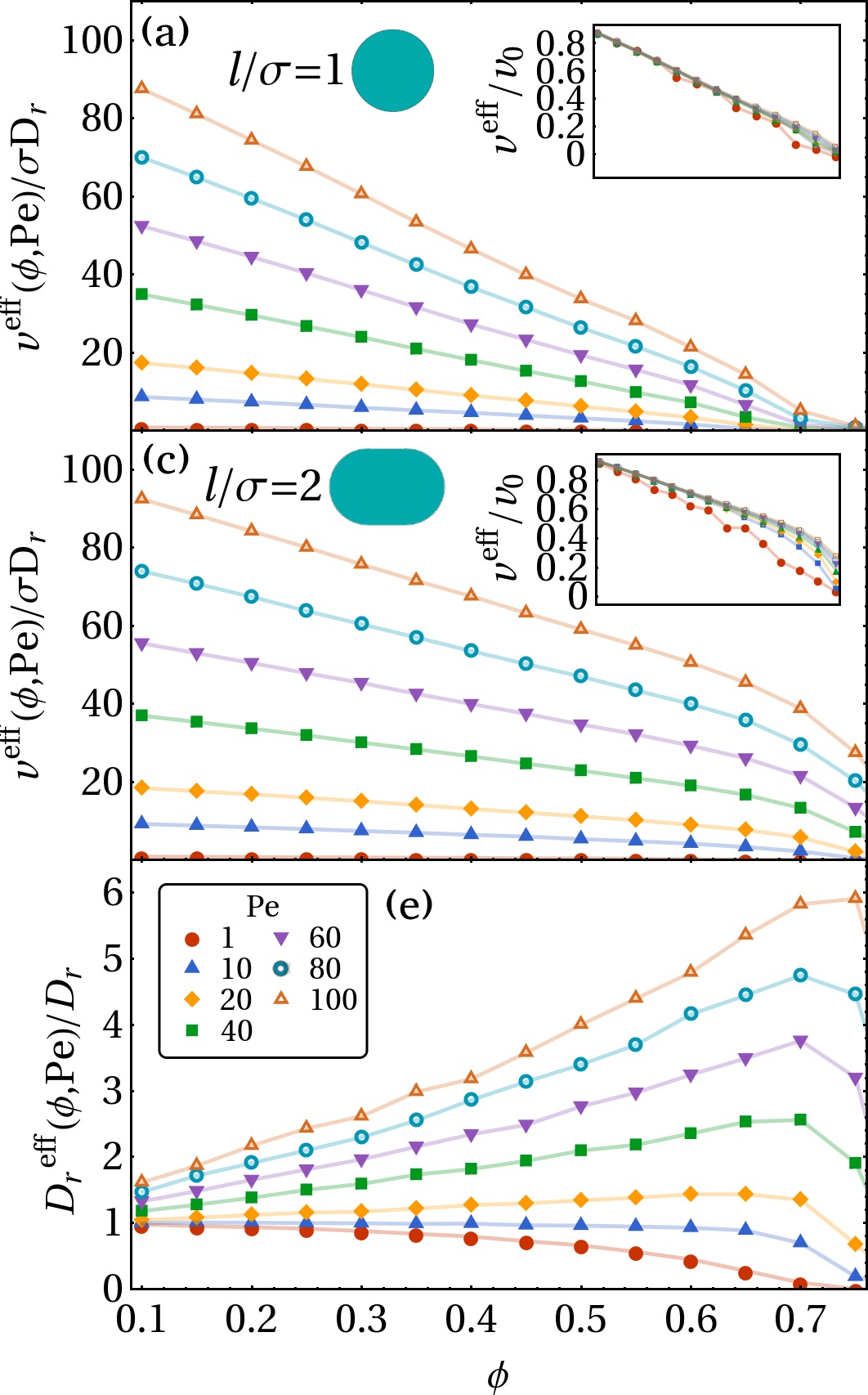}
\qquad
\includegraphics[height=.55\textheight]{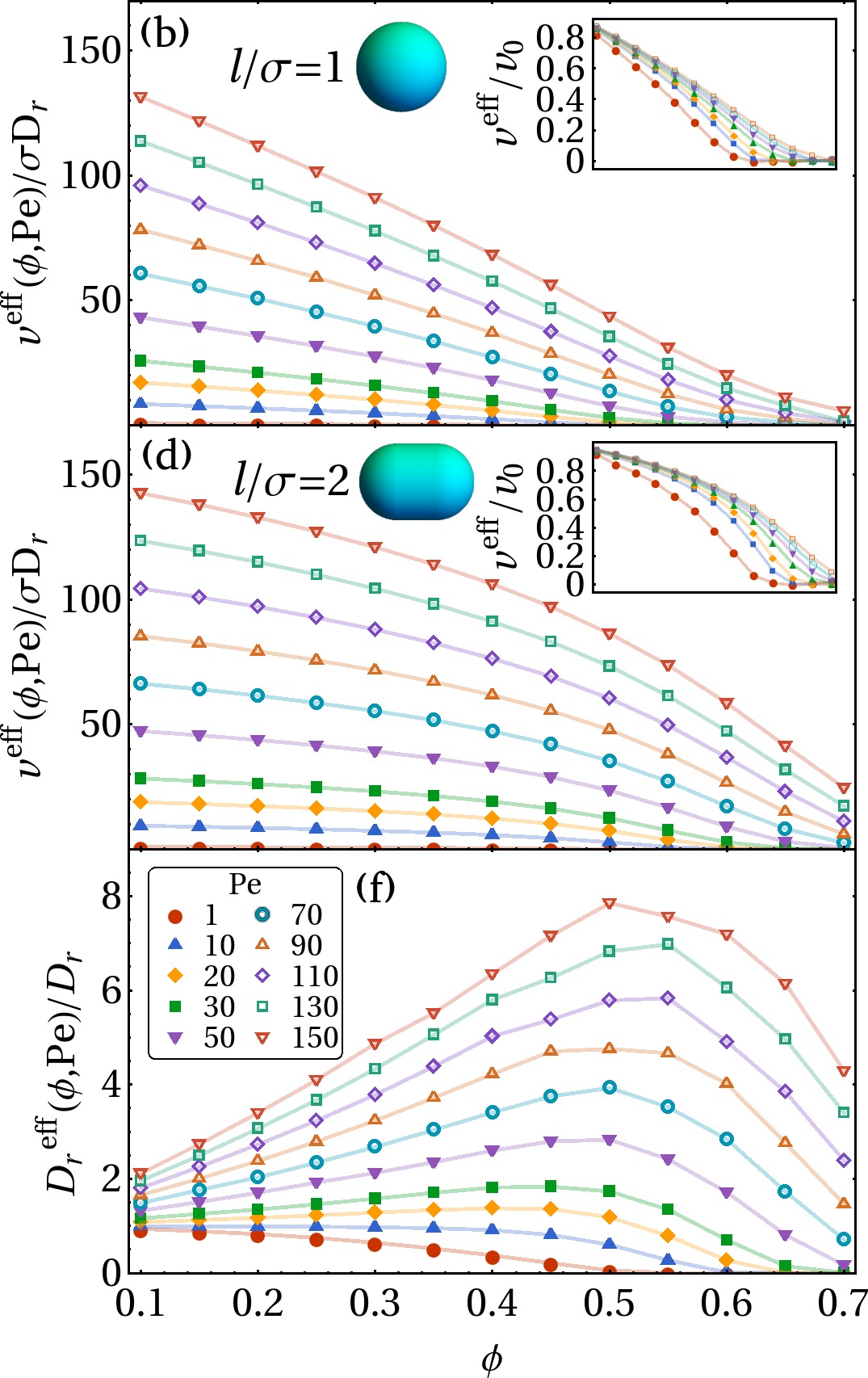}
\caption{\label{fig:EffectiveConstants} Simulation results for the effective self-propulsion speed $v^\text{eff}$ of 2D active disks and 3D active spheres (a,b), effective self-propulsion speed $v^\text{eff}$ for 2D and 3D active rods (c,d), and effective rotational diffusion $D_r^\text{eff}$ for 2D and 3D rods with an aspect ratio $l/\sigma=2.0$ (e,f). All insets show effective velocity divided by input velocity, for  comparison with $v^\text{eff}= v_0(1-\phi/\phi_{cp})$ with $\phi_{cp}$ the close packing density.}
\end{figure*}

Since the main difference between the disk and rod systems is the presence of torque, it is likely that the suppression of MIPS must arise there. In our stability analysis, the only effect of torques is to modify the rotational diffusion. Looking at Eq. (\ref{eq:DensityDiffusionConstant}), it might be possible to suppress MIPS if $D_r^\text{eff}$ is increased enough to make the second term smaller than $D_t^\text{eff}$. Is this the case? Is the rotational diffusion perhaps enhanced so much that we effectively end up with a thermal system again?\\

Not so. Looking at Figs. \ref{fig:EffectiveConstants}e and \ref{fig:EffectiveConstants}f, we can see that $D_r^\text{eff}$ is indeed increased significantly where MIPS vanishes. However, when we insert the actual values of $v^\text{eff}$ and $D_r^\text{eff}$, we see that this is not the case: the typical values of $v^\text{eff}$ are simply too large. So if it is not $D_r^\text{eff}$, it must be $v^\text{eff}$ that contains the key information that allows us to predict MIPS or its suppression. After all, the stability analysis does correctly predict that MIPS is suppressed for high aspect ratios. Comparing the effective swimming speeds of different aspect ratios (Figs. \ref{fig:EffectiveConstants}a and \ref{fig:EffectiveConstants}c, or \ref{fig:EffectiveConstants}b and \ref{fig:EffectiveConstants}d), we see that the rods slow down less with increasing density than the disks. In other words, the rods hinder each other's movement less than the disks do. Why is this? Again we must look to the main difference between the two systems: torque.\\

For disks, one can derive the linear decrease of the velocity with increasing density $v^\text{eff} = v_0 (1-\rho/\rho^*)$ from mean-field theory and kinetic arguments \cite{Stenhammar2013,Fily2014a,Bruss2018}. This is done by assuming that particles slow down at low density due to time spent in binary collisions, which leads to $v^\text{eff}(\rho) \simeq v_0(1-\tau_c/\tau_f)$, where $\tau_c$ is the time spent in a collision and $\tau_f=1/(\sigma v_0 \rho)$ the mean free time between collisions. At low density, we expect the mean free time $\tau_f$ to be mostly unaffected by the presence of torques as long as there are no significant short-range orientational correlations. The duration of collisions $\tau_c$, however, can change significantly when torques are involved. For disks, the duration of their collision---of their hindrance---is determined by how long it takes for them to slide around each other. Rods, however, will rotate to reorient their swimming directions away from the combined center of mass of the collision. This will decrease the collision duration. Since collisions are now shorter, the rods spend more time moving freely: less hindered. Furthermore, this reorientation leads to an enhanced rotational diffusion---exactly what we find.\\

Interestingly, this suggests that an inverse mechanism might also exist. If the torques between two colliding particles cause the particles to rotate towards their center of mass, collisions would be prolonged and MIPS would be enhanced. Precisely this inverse effect was reported earlier in Refs. \cite{Barre2014,Sese-Sansa2018}: MIPS is enhanced for self-propelled particles that align through Vicsek interactions. In binary collisions, the Vicsek torques always rotate particles towards the combined center of mass, increasing the duration of collisions, increasing hindrance and thus enhancing MIPS.\\

\begin{figure}[h]
\includegraphics[width=.95\columnwidth]{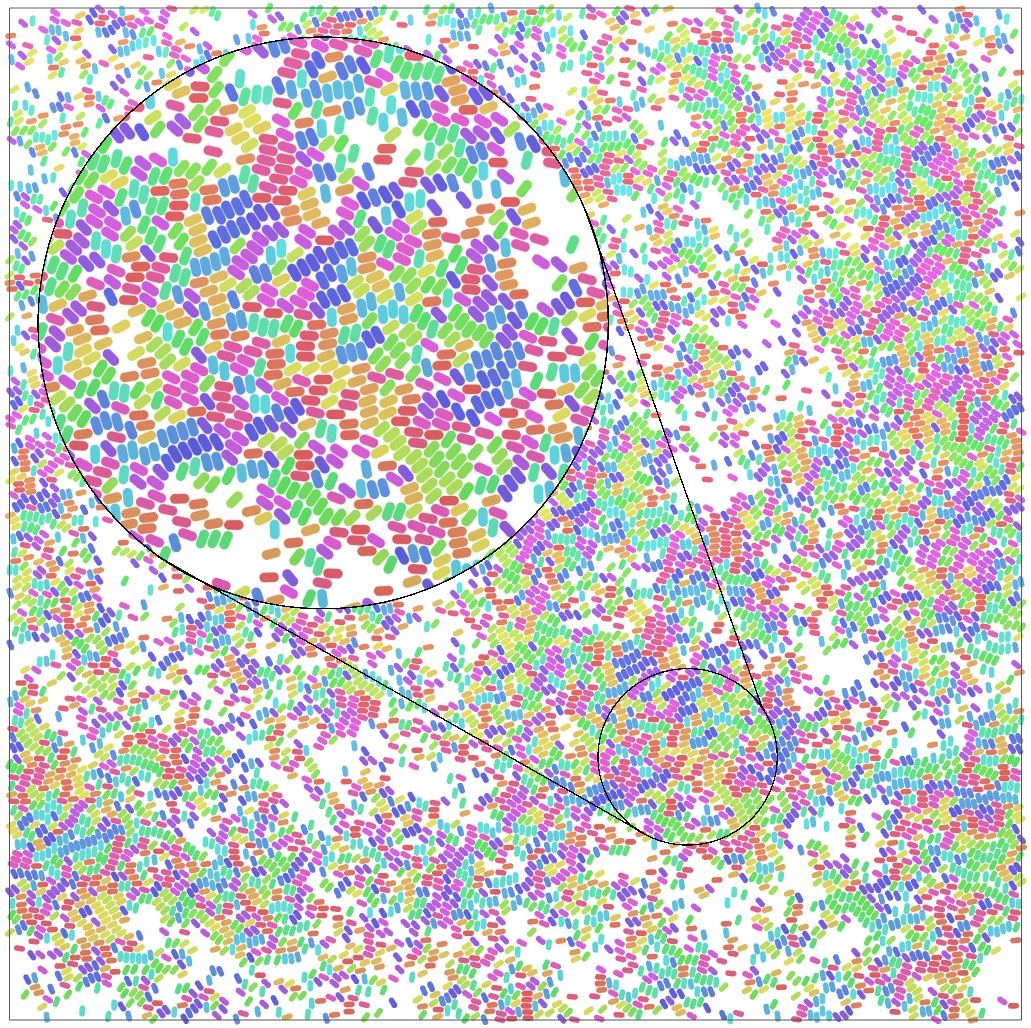}
\caption{\label{fig:2DSnapshots} Simulation snapshot of 2D rods with aspect ratio $l/\sigma=2$ at a packing fraction of $\phi=0.5$ and a P\'eclet number of $\text{Pe}=100$. Particles are coloured according to their orientation. Dense clusters display significant short-range orientational order, and no large-scale phase separation can be seen.}
\end{figure}

Is the changing density dependence of $v^\text{eff}$ with increasing anisotropy enough to completely describe the suppression of MIPS? If we would have a system of self-propelled particles with some arbitrary shape and we would know how the effective swim speed depends on density, could we then predict whether and where it will undergo MIPS? Unfortunately, no. As we can see from the rod phase diagrams in Figs. \ref{fig:2DPhaseDiagram}b and \ref{fig:3DPhaseDiagram}b, our stability analysis predicts the right qualitative trend, but its quantitative prediction is poor. This is probably due to neglecting alignment effects in the stability analysis. As the rod length increases, nematic and polar alignment of the particles start playing a more significant role in their phase behaviour, which is not captured by our theory. For instance consider Fig. \ref{fig:2DSnapshots}, where we show a snapshot of rods at $\phi=0.5$, $\text{Pe}=100$, just outside the MIPS region, and colour particles according to nematic orientation. The clusters formed by the rods have significant short-range nematic order. Incorporating the dynamics of the polarization and nematic fields using theory developed for active nematics \cite{Baskaran2010,Bertin2015} might allow for more accurate predictions for the onset of MIPS for longer rods.

\section{Conclusions \& outlook}
\label{sec:Conclusions}
In this paper, we showed that motility-induced phase separation does not occur for rodlike particles when they become sufficiently anisotropic. This disappearance is observed both from many-particle simulations and from a stability analysis of the homogeneous isotropic phase. The latter provides a simple criterion for the onset of MIPS by considering the effective swimming speed of the particles and their effective rotational diffusion. Both methods agree qualitatively in that MIPS is pushed to higher densities for increasing rod aspect ratio, and they agree quantitatively for short rods that deviate only slightly from disks or spheres. For longer aspect ratios the quantitative agreement is lost, presumable due to alignment interactions that are present, but not taken into account in the stability analysis.\\
We also propose a more intuitive explanation for the suppression mechanism. MIPS relies on particles slowing down sufficiently with increasing density \cite{Cates2015}. This hindrance is closely linked to the duration of collisions between particles \cite{Stenhammar2013,Fily2014a,Bruss2018}. Excluding torques, the duration of collisions is determined by how long it takes for them to slide along one another. Including torques can dramatically decrease the duration of collisions by rotating the forward axes of the self-propelled particles away from each other. Formulated in this way, we can also explain the results of Refs. \cite{Barre2014,Sese-Sansa2018}, where MIPS is enhanced for particles with Vicsek interactions. Simply put, Vicsek torques prolong particle collisions, while rodlike excluded volume torques shorten them. Intriguingly, this provides us with a particle design tool to enhance or suppress MIPS. MIPS is enhanced for Vicsek-like interactions \cite{Barre2014,Sese-Sansa2018}, for faceted, concave and/or rough particles \cite{Ilse2016,Prymidis2016,Cugliandolo2017}, while it is suppressed for smooth particles and rodlike shapes \cite{Shi2018}. In addition to steric interactions, hydrodynamic interactions between active particles also play an important role in whether or not MIPS can form. While hydrodynamics seems to usually suppress MIPS \cite{Zottl2014,Matas-Navarro2014,Blaschke2016}, the details depend on whether particles are ``pushers" or ``pullers" and on the dimensionality \cite{Oyama2016,Oyama2017}.\\
Despite recent advancements, the role of torque in active systems is still not well understood. Much of the developed theory has been restricted to the torque-free regime, but recent numerical studies suggest that torque, either from boundaries \cite{Solon2015a} or from particle interactions \cite{Barre2014,Sese-Sansa2018}, can have a significant effect on the structure and dynamics of active matter systems. In order to understand active matter beyond torque-free model systems, more theoretical work is needed to elucidate the influence of torques in active systems.\\

\section*{Supplementary Material}
\label{sec:SupplementaryMaterial} See supplementary material for an analysis of the significance of the effective translational diffusion $D_t^\text{eff}$, the hydrodynamic friction coefficients for short spherocylinders, a finite-size analysis of the effective self-propulsion velocity $v^\text{eff}$ and effective rotational diffusion $D_r^\text{eff}$, and additional simulation snapshots.

\section*{Acknowledgements}
\label{sec:Acknowledgements} This work is part of the D-ITP consortium, a program of the Netherlands Organisation for Scientific Research (NWO) that is funded by the Dutch Ministry of Education, Culture and Science (OCW). We acknowledge financial support from an NWO-VICI grant. We would like to acknowledge Bram Bet for useful discussions and information on the friction coefficients of short spherocylinders.

\appendix*
\section{Stability analysis of the homogeneous isotropic phase, including torque}
\label{sec:Approximations}

There is currently no general theory for out-of-equilibrium statistical physics from which we can derive the onset of MIPS. In equilibrium, one can derive spinodal and binodal lines by considering (derivatives of) the free energy. Out of equilibrium, effective free energies can only be constructed under specific circumstances \cite{Cates2015}. Nevertheless, there are other ways to derive criteria for the onset of MIPS. We are aware of three ways to obtain such criteria: by constructing an effective free energy and proceeding as in equilibrium \cite{Cates2013,Cates2015}, by looking at the particle flux balance between a dense cluster and a dilute gas phase \cite{Redner2013,Redner2013a}, and by a stability analysis of density fluctuations of the homogeneous isotropic phase \cite{Bialke2013,Speck2015}. The first method cannot be applied directly to our system, as one of its underlying assumptions is that no torques act between the particles. The second method is also likely to fail for rods, as it relies on the assumption that the orientations of particles in the boundary of the dense cluster evolve diffusively. Thus, we opt for the third method: deriving a criterion for the (in)stability of the homogeneous isotropic phase to density fluctuations, by extending the mean-field-like method from Ref. \cite{Speck2015} to 3D systems with torques. This is done by capturing the effect of the torques into a modified, \textit{effective} rotational diffusion $D_r^\text{eff}$. Here we describe this extended derivation.

\subsection{Effective Smoluchowski equation}
To render the problem analytically tractable, our first goal is to simplify the effect of the pairwise forces and torques. We will do this using a mean-field-like approximation. Following the same procedure as Refs. \cite{Bialke2013,Speck2015}, we start from the Smoluchowski equation for the one-particle probability density $\psi(\bm{r},\bm{\hat{e}},t)$, given by
\begin{align}
\label{eq:FullSmoluchowski}
\partial_t \psi = & -\bm{\nabla} \cdot \left( v_0 \bm{\hat{e}} \psi + \beta D_t \bm{\mathcal{F}} - D_t \bm{\nabla} \psi \right) \\
& - \bm{\mathcal{R}} \cdot \left( \beta D_r \bm{\mathcal{T}} - D_r \bm{\mathcal{R}} \psi \right) \nonumber,
\end{align}
where $\bm{\nabla}$ are the 2D and 3D gradient operators and $\bm{\mathcal{R}}$ is the rotation operator, defined as $\mathcal{R}=\partial_\theta$ in 2D and $\bm{\mathcal{R}}=\bm{\hat{e}}\times\bm{\nabla}_{\bm{\hat{e}}}$ in 3D. Note that similar to our numerical model we neglect the influence of particle shape anisotropy on the translational diffusion and simply set $\bm{D}_t=D_t \bm{\mathcal{I}}$. The pairwise force density $\bm{\mathcal{F}}$ and torque density $\bm{\mathcal{T}}$, which arise due to the particle-particle interactions of a pair potential $V_{\bm{\hat{e}}_1,\bm{\hat{e}}_2}(\bm{r}_1,\bm{r}_2)$, can then be written in terms of the two-body probability density  $\psi^{(2)}_{\bm{\hat{e}}_1,\bm{\hat{e}}_2}(\bm{r}_1,\bm{r}_2,t)$ as
\begin{align}
& \bm{\mathcal{F}}(\bm{r}_1,\bm{\hat{e}}_1,t) \equiv \label{eq:Fpsi2} \\ 
& \int \text{d}\bm{r}_2\text{d}\bm{\hat{e}}_2\left(-\nabla_1V_{\bm{\hat{e}}_1,\bm{\hat{e}}_2}(\bm{r}_1,\bm{r}_2)\right) \psi^{(2)}_{\bm{\hat{e}}_1,\bm{\hat{e}}_2}(\bm{r}_1,\bm{r}_2,t) \nonumber;
\end{align}
\begin{align}
& \bm{\mathcal{T}}(\bm{r}_1,\bm{\hat{e}}_1,t) \equiv \\ 
&\int \text{d}\bm{r}_2\text{d}\bm{\hat{e}}_2\left(-\mathcal{R}_1V_{\bm{\hat{e}}_1,\bm{\hat{e}}_2}(\bm{r}_1,\bm{r}_2)\right) \psi^{(2)}_{\bm{\hat{e}}_1,\bm{\hat{e}}_2}(\bm{r}_1,\bm{r}_2,t). \nonumber
\end{align}
In order to close this hierarchy, the force and torque densities $\bm{\mathcal{F}}$ and $\bm{\mathcal{T}}$ need to be expressed in terms of the one-body PDF. To do so, we first use the identity
\begin{equation}
\psi^{(2)}_{\bm{\hat{e}}_1\bm{\hat{e}}_2}(\bm{r}_1,\bm{r}_2,t)=\psi(\bm{r}_1,\bm{\hat{e}}_1,t)\psi(\bm{r}_2,\bm{\hat{e}}_2,t)g_{\bm{\hat{e}}_1\bm{\hat{e}}_2}(\bm{r}_1,\bm{r}_2,t)
\end{equation}
to rewrite Eq. (\ref{eq:Fpsi2}) as $\bm{\mathcal{F}}=\bm{\tilde{\mathcal{F}}} \psi$, where
\begin{align}
& \bm{\tilde{\mathcal{F}}}(\bm{r}_1,\bm{\hat{e}}_1,t) \equiv \label{eq:Ftildepsi} \\
& \int \text{d}\bm{r}_2\text{d}\bm{\hat{e}}_2\left(-\nabla_1V_{\bm{\hat{e}}_1,\bm{\hat{e}}_2}(\bm{r}_1,\bm{r}_2)\right) \psi(\bm{r}_2,\bm{\hat{e}}_2,t) g_{\bm{\hat{e}}_1,\bm{\hat{e}}_2}(\bm{r}_1,\bm{r}_2,t) \nonumber.
\end{align}
To obtain a closure, we make the following assumptions. First, we assume that the force $\bm{\mathcal{F}}$ acts along the direction of self-propulsion, i.e. $\bm{\mathcal{F}}=(\bm{\mathcal{F}}\cdot\bm{\hat{e}})\bm{\hat{e}}$. Whereas this is exact in a homogeneous, isotropic bulk as dictated by symmetry, in general we neglect a possible second component that is perpendicular to $\bm{\hat{e}}$. In Ref. \cite{Speck2015}, Speck et al. consider this second component to act along the gradient of the one-particle PDF i.e. $\bm{\mathcal{F}}=(\bm{\mathcal{F}}\cdot\bm{\hat{e}})\bm{\hat{e}}+a\nabla \psi$. This additional assumption leads to a modified translational diffusion $D_t^\text{eff}=(1-\beta a)D_t$. We measured the magnitude of this modification for 3D spheres and rods, and found that the modification provided by $\beta a$ is of negligible influence on the location of the phase boundaries. Thus, we do not consider this additional component here and simply set $a=0$, i.e. $D_t^\text{eff}=D_t$, from now on. We did not explicitly check the validity of this assumption in the 2D case, but see no reason to assume a difference.\\
To continue our derivation, we make the second assumption that $\bm{\tilde{\mathcal{F}}}\cdot\bm{\hat{e}}$ is linear in the local density $\rho(\bm{r},t)=\int\text{d}\bm{\hat{e}}\psi(\bm{r},\bm{\hat{e}},t)$ and has no further dependence on $(\bm{r},\bm{\hat{e}},t)$:
\begin{equation}
\bm{\tilde{\mathcal{F}}}(\bm{r},\bm{\hat{e}},t)\cdot\bm{\hat{e}} = - \rho(\bm{r},t)\zeta(\bar{\rho},v_0).
\end{equation}
Here the constant $\zeta$ is independent of $(\bm{r},\bm{\hat{e}},t)$, but can still depend on the mean density $\bar{\rho}=N/A$ (or $N/V$ in 3D) and the self-propulsion strength $v_0$. In this way, using Eq. (\ref{eq:FullSmoluchowski}), the effect of the interaction forces can be absorbed into a modified self-propulsion velocity $v^\text{eff}$, which is given by
\begin{equation}
v^\text{eff} = v_0 - \beta D_t \rho(\bm{r},t)\zeta(\bar{\rho},v_0).
\end{equation}
For the torques, we make the approximation that its only influence is to modify the rotational diffusion i.e.
\begin{equation}
\bm{\mathcal{T}}(\bm{r},\bm{\hat{e}},t) \approx b \bm{\mathcal{R}}_1\psi(\bm{r},\bm{\hat{e}},t).
\end{equation}
Proceeding on, we assume the corresponding constant to be homogeneous and isotropic, depending only on the mean density and self-propulsion: $b=b(\bar{\rho},v_0)$. With these approximations we can simplify Eq. (\ref{eq:FullSmoluchowski}) as the Smoluchowski equation for an active ideal gas:
\begin{equation}
\label{eq:IdealGasSmoluchowski}
\partial_t \psi =  -\bm{\nabla} \cdot \left( v^\text{eff} \bm{\hat{e}} \psi -  D_t \bm{\nabla} \psi \right) + D_r^\text{eff} \bm{\mathcal{R}}\cdot\bm{\mathcal{R}} \psi,
\end{equation}
where $v^\text{eff}$ and $D_r^\text{eff}$ are now the \textit{effective} self-propulsion and rotational diffusion constant, respectively.

\subsection{Stability analysis of the homogeneous isotropic phase}
\label{sec:StabilityAnalysis}

Now that we have reduced the full Smoluchowski Eq. (\ref{eq:FullSmoluchowski}) into the ideal-gas form of Eq. (\ref{eq:IdealGasSmoluchowski}), we can perform a linear stability analysis on the homogeneous isotropic phase. We start by defining the relevant moments of the one-particle PDF $\psi(\bm{r},\bm{\hat{e}},t)$,\\
\begin{align}
\rho(\bm{r},t) &= \int\text{d}\bm{\hat{e}}\psi(\bm{r},\bm{\hat{e}},t) \quad (\text{density}); \\
m_\alpha(\bm{r},t) &= \int\text{d}\bm{\hat{e}}e_\alpha\psi(\bm{r},\bm{\hat{e}},t) \quad (\text{polarization}); \\
S_{\alpha\beta}(\bm{r},t) &= \int\text{d}\bm{\hat{e}}(e_\alpha e_\beta - \frac{1}{d}\delta_{\alpha\beta})\psi(\bm{r},\bm{\hat{e}},t)\quad (\text{nematic}).
\end{align}
Here, the Greek indices label the Cartesian vector- or tensor components, and in the following we shall employ the Einstein summation convention. Considering the same moments of the ideal gas Smoluchowski equation (\ref{eq:IdealGasSmoluchowski}) yields the following evolution equations:
\begin{align}
\partial_t \rho = \label{eq:dtRho}
& - \bm{\nabla} \cdot \left( v^\text{eff} \bm{m} - D_t\bm{\nabla} \rho \right); \\
\partial_t m_\alpha = \label{eq:dtM}
& - \partial_\beta [ v^\text{eff}(S_{\alpha\beta} + \frac{1}{d}\rho\delta_{\alpha\beta}) - D_t \partial_\beta m_\alpha ] \\
& - (d-1) D_r^\text{eff} m_\alpha; \nonumber \\
\partial_t S_{\alpha\beta} = \label{eq:dtS}
& - \partial_\gamma [ v^\text{eff}(B_{\alpha\beta\gamma} - \frac{1}{d}m_\gamma \delta_{\alpha\beta\gamma}) - D_t\partial_\gamma S_{\alpha\beta} ] \\ 
& - d(d-1) D_r^\text{eff}S_{\alpha\beta}. \nonumber
\end{align}
Here $B$ is the next (third) order moment. The structure of this hierarchy of time-evolution equations (\ref{eq:dtRho})-(\ref{eq:dtS}) is such that the time-derivative of each moment depends linearly on itself and lower order ones, and on the next one. However, as we shall see, moments beyond $m$ are irrelevant for the instability we wish to consider.\\
A steady-state solution to Eq. (\ref{eq:IdealGasSmoluchowski}) is the homogeneous isotropic phase: $\psi(\bm{r},\bm{\hat{e}},t)\propto\bar{\rho}$. Expressed in terms of the moment equations (\ref{eq:dtRho})-(\ref{eq:dtS}), this gives $\rho(\bm{r},t)=\bar{\rho}$ and $\bm{m}(\bm{r},t)=\bm{S}(\bm{r},t)=0$. To obtain a criterion for the stability of this solution, we investigate the behaviour of small perturbations to the homogeneous state:
\begin{align} 
\rho(\bm{r},t) &= \bar{\rho}+\delta\rho(\bm{r},t);
\\ \bm{m}(\bm{r},t) &= \bm{\delta m}(\bm{r},t);
\\ \bm{S}(\bm{r},t) &= \bm{\delta S}(\bm{r},t).
\end{align}
Since MIPS is a macroscopic phase separation, we should study the instability with respect to long-range perturbations i.e. perturbations with small spatial gradients. In this limit, the dynamics are dominated by the terms in Eqs. (\ref{eq:dtRho})-(\ref{eq:dtS}) with the fewest gradients. Of the three moments, it is $\rho$ whose time evolution is slowest. Its timescale is of order $\nabla^{-1}$, while $\bm{m}$ and $\bm{S}$ evolve as $(D_r^\text{eff})^{-1}\sim\nabla^0$. As we are interested in the evolution of the density perturbations, i.e. of the slow variable, we can assume that at any given time, the higher moments $\bm{m}$ and $\bm{S}$ are given by their steady-state solutions that correspond to the density profile $\rho(\bm{r},t)$ at that instant. Solving Eq. (\ref{eq:dtS}) for its steady-state solution $\delta S_{\alpha\beta}$ reveals that it scales as $\mathcal{O}(\nabla^1)$. Therefore, its contribution to the evolution of polarization perturbations (Eq. (\ref{eq:dtM})) is of higher order. To leading order, the evolution of polarization perturbations is then given by
\begin{equation}
\bm{\delta m}(\bm{r},t) = - \frac{1}{d(d-1)D_r^\text{eff}} \bm{\nabla} \left(v^\text{eff}(\bm{r},t)\rho(\bm{r},t)\right).
\end{equation}
Recalling that $v^\text{eff}=v_0-\beta D_t \rho(\bm{r},t)\zeta(\bar{\rho},v_0)$, we can take this gradient explicitly and obtain
\begin{align}
\bm{\delta m}(\bm{r},t) = - \frac{1}{d(d-1)D_r^\text{eff}} \left(v_0-2\beta D_t \rho(\bm{r},t)\zeta\right)\bm{\nabla}\rho(\bm{r},t).
\end{align}
Using this result, the equation for the time evolution of density perturbations becomes
\begin{equation}
\partial_t \delta \rho (\bm{r},t) = \mathcal{D}_{\delta\rho}(\bar{\rho},v_0)\bm{\nabla}^2\delta\rho(\bm{r},t),
\label{eq:DensityDiffusion}
\end{equation}
which is a diffusion equation with diffusion constant
\begin{align}
\mathcal{D}_{\delta\rho}(\bar{\rho},v_0) & = D_t + \frac{(v_0-\beta D_t \bar{\rho}\zeta)(v_0-2\beta D_t \bar{\rho}\zeta)}{d(d-1)D_r^\text{eff}} \label{eq:DensityDiffusionConstant} \\
& = D_t + \frac{v^\text{eff}(2v^\text{eff}-v_0)}{d(d-1)D_r^\text{eff}} \nonumber.
\end{align}
Whenever the diffusion constant $\mathcal{D}_{\delta\rho}$ is negative, density perturbations $\delta\rho(\bm{r},t)$ will grow. Therefore, the region in $(\bar{\rho},v_0)$-space where the homogeneous isotropic phase becomes unstable is then given by the condition $\mathcal{D}_{\delta\rho}(\bar{\rho},v_0)<0$. This can only occur for self-propulsion velocities $v_0$ above the critical threshold $v^*=2\sqrt{2}\sqrt{d(d-1)}\sqrt{D_t D_r^\text{eff}}$.

%

\end{document}